\documentclass[aps,twocolumn,showpacs,groupedaddress,floatfix]{revtex4}

\usepackage{graphicx}
\usepackage{subfigure}
\usepackage{subfloat}
\usepackage{amssymb}
\usepackage{amsmath}
\usepackage{color}
\usepackage{comment}
\usepackage{cleveref}

\newcommand{\vect}[1]{{\mathbf #1}}

\begin{document}


\title{Enlarging and cooling the N\'{e}el state in an optical lattice}

\author{Charles J. M. Mathy}
\affiliation{ITAMP, Harvard-Smithsonian Center for Astrophysics, Cambridge, Massachusetts 02138, USA}
\affiliation{Physics Department, Harvard University, Cambridge, Massachusetts 02138, USA}

\author{David A. Huse}
\affiliation{Physics Department, Princeton University, Princeton, New Jersey 08544, USA}

\author{Randall G. Hulet}
\affiliation{Department of Physics and Astronomy, Rice University, Houston, Texas 77005, USA} %

\date{\today}
 \begin{abstract}
We propose an experimental scheme to favor both the realization and the detection of the N\'{e}el state in a two-component gas of
ultracold fermions in a three-dimensional simple-cubic optical lattice.
By adding three compensating Gaussian laser beams to the standard three pairs of retroreflected lattice beams,
and adjusting the relative waists and intensities of the beams,
one can significantly enhance the size of the N\'{e}el state in the trap, thus increasing the signal of optical Bragg scattering.
Furthermore, the additional beams provide for adjustment of the local chemical potential and the possibility to evaporatively cool the gas while in the lattice. Our proposals are also relevant to efforts to realize other many-body quantum phases in optical lattices.
\end{abstract}



%

\maketitle

\subsection{Introduction}

Cold atom experiments provide a uniquely versatile platform for realizing and probing strongly correlated quantum phases of matter. However, no experiment to date has measured a phase in an optical lattice whose ordering is set by a magnetic scale such as the superexchange energy. Experiments have realized Mott insulators of both bosons \cite{Greiner02} and fermions \cite{Jordens08,Schneider08}, but the temperatures achieved are higher than those required for magnetic ordering \cite{Paiva10,Jordens10,Fuchs11,DeLeo10}. Furthermore,
the lack of a heat bath imposes restrictions on experimental schemes \cite{Ma08,Bernier09,Zhou09,Cone12}, in which the entropy must be pushed out from the center of the trap where the phase of interest is realized.
In the case of gapped phases, equilibration is impeded by the long timescales for heat and mass transport \cite{Hung08, Parish09}.
Finally, an experimental setup should strive to have the phase of interest occupy as large a region of the trap as possible to enhance the detectability of the ordering, for example, by Bragg scattering of light \cite{Corcovilos10}.

In this work, we propose an all-optical scheme that
addresses these issues for the N\'{e}el phase of ultracold fermions in a simple-cubic optical lattice, and discuss its relevance to other efforts to realize strongly-correlated many-body quantum phases. The objective is to maximize the size of the phase of interest in the trap to enhance the Bragg signal, realize the N\'eel phase in the region of parameter space that was previously calculated to have maximal superexchange interactions \cite{Mathy2009}, and provide a setup that will allow for cooling when the center of the trap becomes a Mott insulator, for which heat and mass transport are inhibited.
We show that these objectives can all be met simply by introducing three compensating laser beams on top of the three retroreflected lattice beams.
These compensating beams have different Gaussian beam waists than the lattice beams and are oppositely detuned,
so that if the lattice beams generate an attractive potential for the atoms, the compensating beams are repulsive, and vice-versa.
The compensating beams 
allow the overall chemical potential of the system to stay in the gap of the phase of interest over a larger fraction of the cloud.  We propose to do this in a
different manner than was analyzed in Ref. \cite{Ma08}, where the trap potential was flattened by making it either quartic or something close to a square well.
In both of these cases the walls of the trap are made steeper when the bottom is made flatter, and as a result the number of atoms in the outer gapless part of
the cloud is reduced, making the system very sensitive to variations in the total atom number.
In our setup, the confining potential is smooth and decays as a Gaussian at large distances.  The trap is filled so that there remain many atoms in the outer
gapless ``reservoir" parts of the cloud, as discussed below.  Thus the system will not be sensitive to small variations in the total atom number.
Furthermore, our setup allows for direct evaporative cooling of the system while in the lattice.

\begin{figure}[htp]
\centering
\parbox{1\linewidth} {
\subfigure{ \includegraphics[width=0.9\linewidth, clip=true,trim= 0 0 0 00]{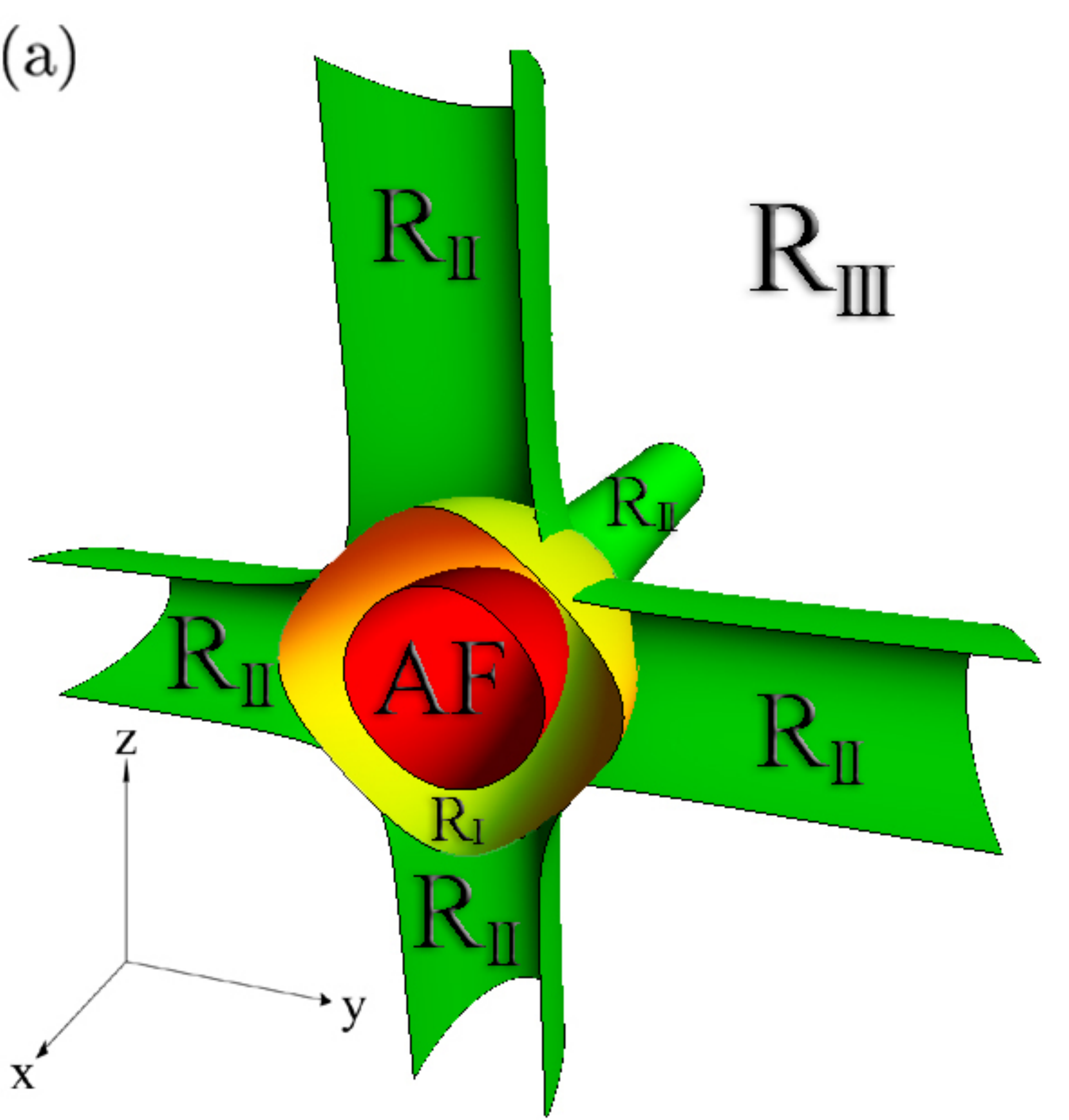}
\label{fig:Phases}
}
}
\qquad
\parbox{1\linewidth}
{
\subfigure{
\includegraphics[width=1\linewidth, clip=true,trim= 0 0 0 0]{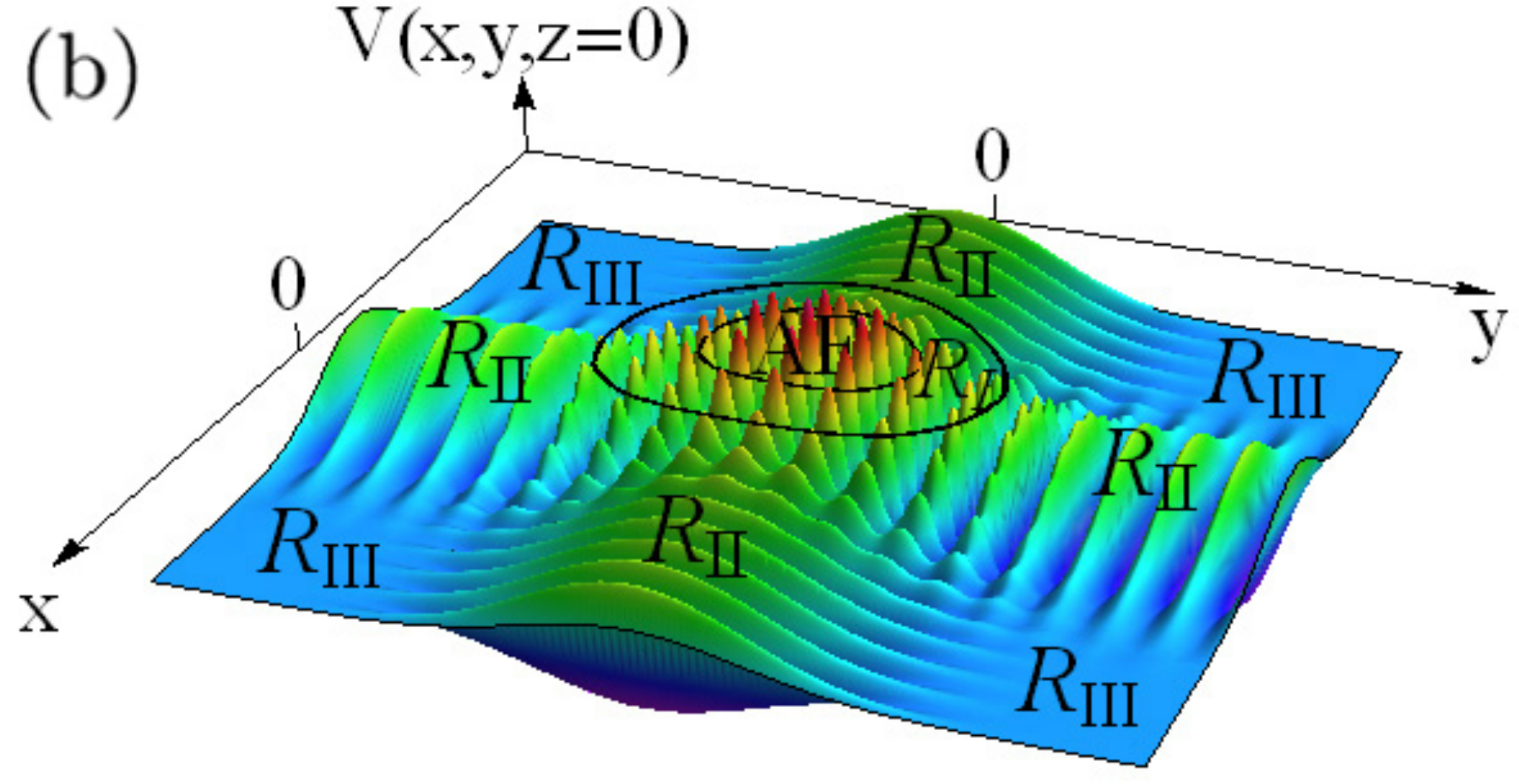}
\label{fig:LaserCut}
}
}
\caption{(Color online) Phases and potentials in the trap. \subref{fig:Phases}  Schematic representation of the distribution of phases. The three lattice beams run along the $x$-, $y$- and $z$-axes. They meet and form a simple-cubic lattice at the center.  The antiferromagnetic (AF) Mott-insulating phase with filling of one atom per lattice site is found here.
Surrounding the AF phase in the region where all three lattice beams are non-negligible is a paramagnetic Fermi gas phase $R_I$ (``Reservoir I").
The regions where only one of three lattice beams has significant intensity are denoted
$R_{II}$.  The region where all lattice beams are negligible make up $R_{III}$.  Atoms can be contained in $R_{III}$ by an additional confining $V_{ext}(\vect{r})$.
\subref{fig:LaserCut} Total potential at $z=0$: $V(x,y,z=0)$. Compensating beams, of optical potential opposite to the lattice beams, run along the $x$-, $y$- and $z$- axes to maximize the size of the AF region.
}
\label{fig:LaserPhases}
\end{figure}

\begin{figure}[t!]
\includegraphics[width=1\linewidth]{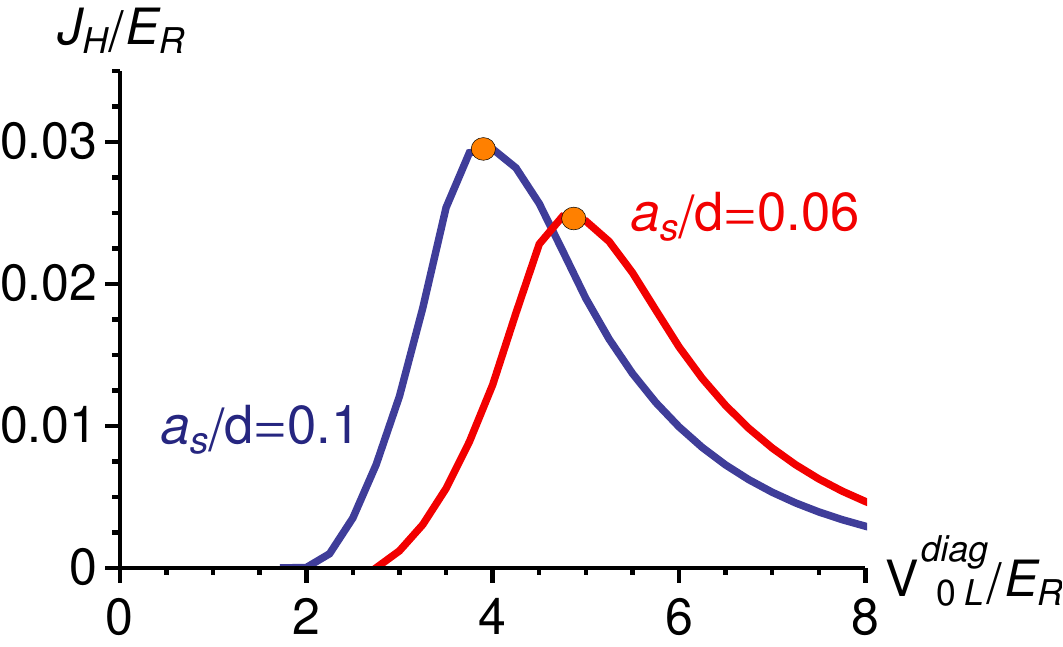}
\caption{ (Color online)
Plot of the Hartree estimate of the effective antifferomagnetic exchange $J_H$ in a simple-cubic optical lattice with lattice depth $V_{0L}^{diag}$, for interaction strengths $a_s=0.1d$ (blue) and $a_s=0.06d$ (red) where d is the lattice spacing.
$J_H$ is twice the difference in energy per bond between the antiferromagnetic and ferromagnetic states, obtained in the Hartree approximation (see \cite{Mathy2009} for details). A large $J_H$ leads to a large ordering temperature, and fast time scales for equilibration and entropy transport. The highlighted points (orange) correspond to the maximum of $J_H$ as a function of $V_0$ for a given interaction strength $a_s/d$. Employing the local-density approximation, the trap shown in Fig. \ref{fig:LaserPhases} is simple-cubic along the diagonal directions with lattice depth $V_{0L}^{diag}$ decreasing as one leaves the trap center. Thus by arranging to have $V_{0L}^{diag}$ be close to the value that maximizes $J_H$ over a large portion of the trap, the conditions for realizing the N\'eel phase are optimized.
}
\label{fig:JH}
\end{figure}


\begin{figure*}[htp]
 \includegraphics[width=1\linewidth, clip=true,trim= 0 0 0 00]{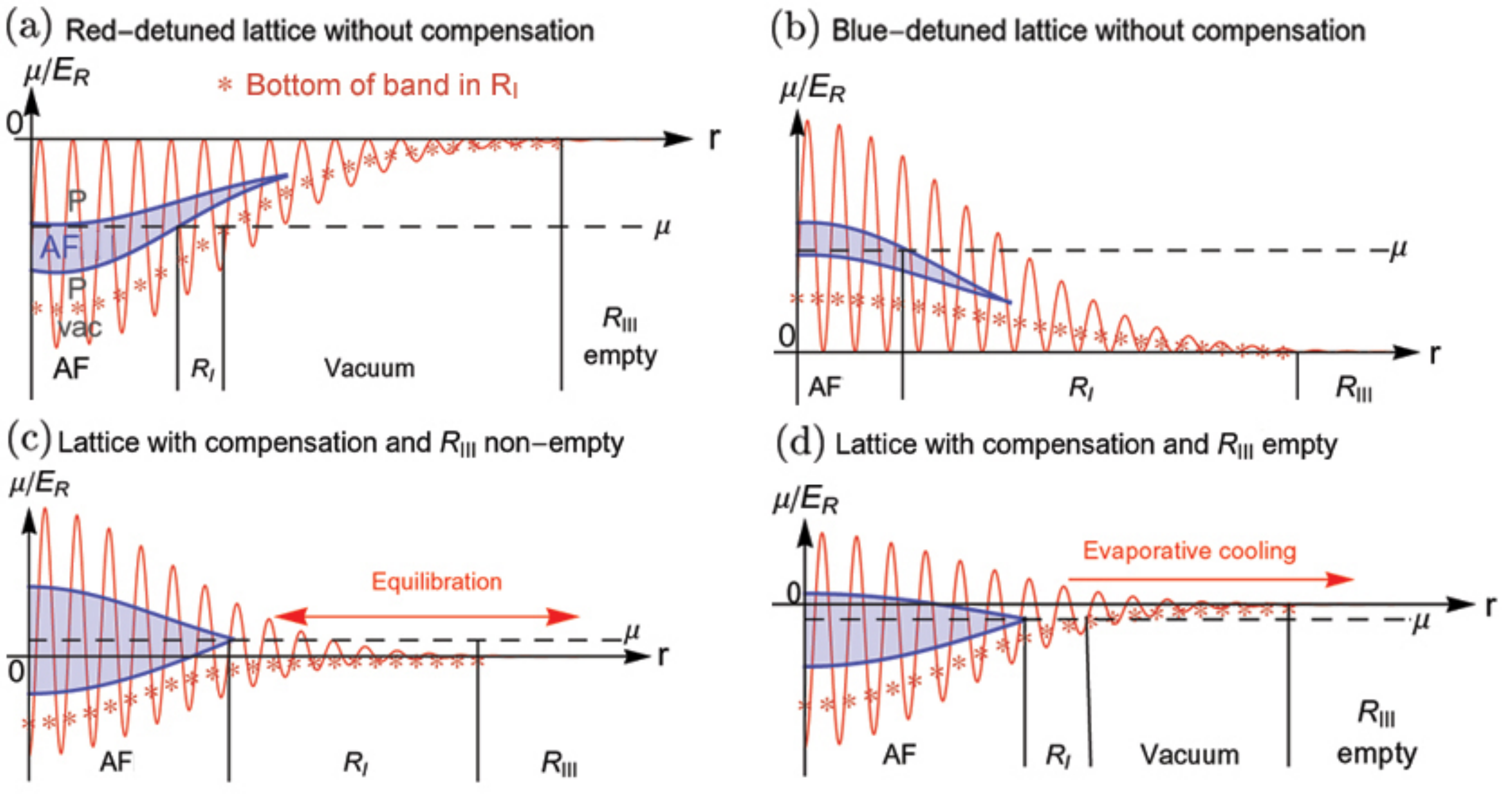}
\caption{(Color online) Phases along the diagonal direction (schematic). The horizontal axes denotes the radial coordinate $r=\sqrt{x^2+y^2+z^2}$. The red lines are schematic depictions of the potential along the diagonals: $V(\pm x,\pm y,\pm z)$. The dashed lines indicate the value of the overall chemical potential $\mu$.
The red stars denote the lowest state of the system, within a local density approximation. If $\mu$ lies below the red stars, the system is in the vacuum. The shaded blue regions correspond to the antiferromagnetic Mott insulator phase (AF). For $\mu$ above the red stars but outside the shaded blue region, the gas is a paramagnet, which we call $R_I$ where the lattice potential is non-negligible.
$R_{III}$ is the region where all lattice potentials are negligible and is occupied only if $\mu >0$. There are four general classifications: (a) attractive lattice without compensation; (b) repulsive lattice without compensation; (c) attractive or repulsive lattice with compensation and $R_{III}$ is non-empty; (d) attractive or repulsive lattice with compensation and $R_{III}$ is empty. 
These classifications are depicted more quantitatively in Figures \ref{fig:musnocomp} to \ref{fig:musa0p06}. 
With compensation, the AF phase has its largest possible size when $\mu$ is set so it coincides with the point where the Hubbard gap closes; this case is shown here. In the case of (d), $\mu$ is chosen to be just below the zero of energy. This situation enables evaporative cooling, as particles with energies above zero correspond to excitations and are able to leave the trap. In (c), the system is in thermal contact with the outside reservoir $R_{III}$. The temperature of the system in equilibrium is set by the temperature of $R_{III}$, which itself may be evaporatively cooled during the experiment.
}
\label{fig:Potentials}
\end{figure*}

\begin{figure}[htp]
\centering
\parbox{1\linewidth} {
\subfigure{ \includegraphics[width=0.95\linewidth, clip=true,trim= 0 0 0 00]{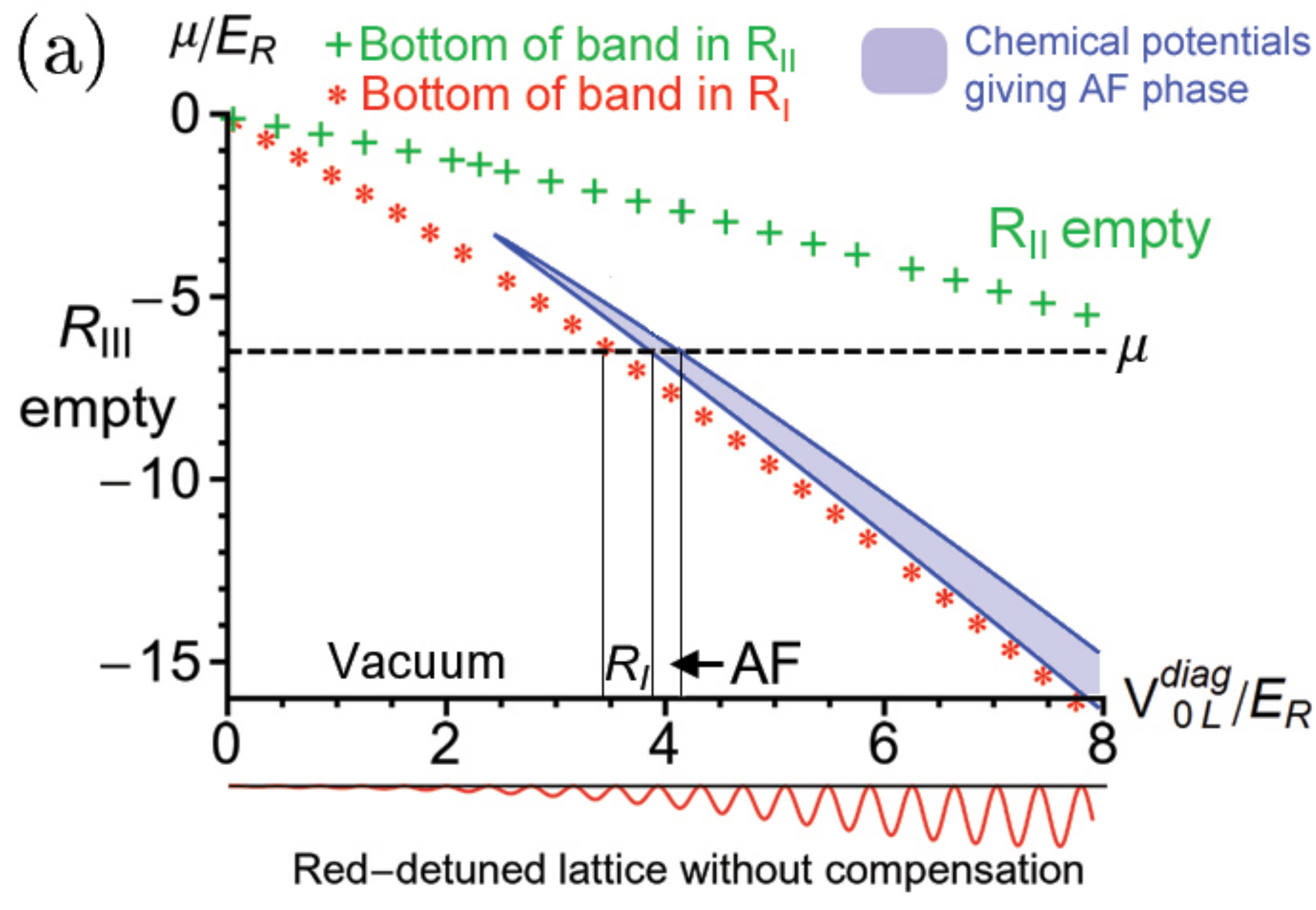}
\label{fig:musrednocomp}
}
}
\qquad
\parbox{1\linewidth}
{
\subfigure{
\includegraphics[width=1\linewidth, clip=true,trim= 0 0 0 0]{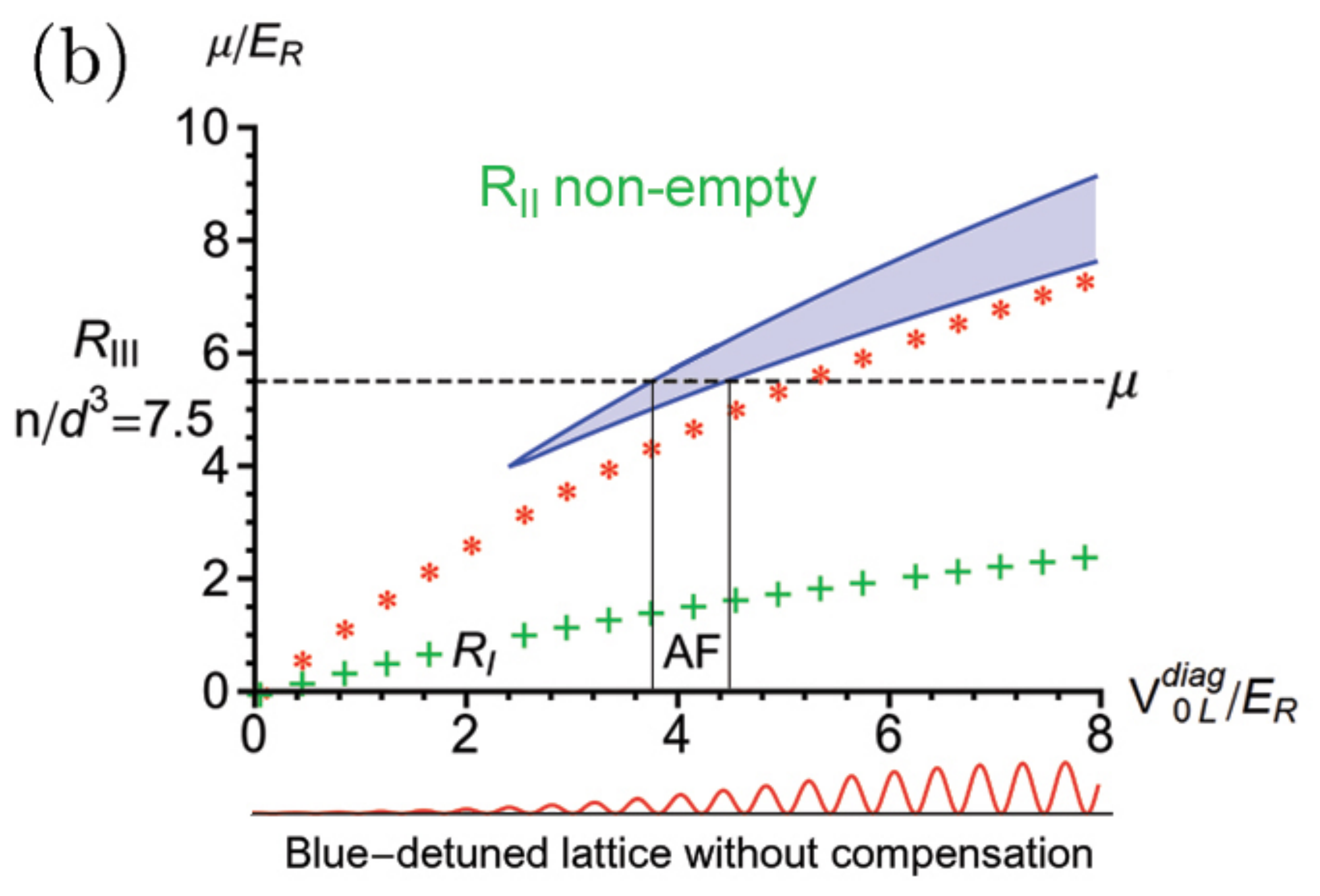}
\label{fig:musbluenocomp}
}
}
\caption{(Color online)
Phases without compensation and with $a_s=0.1d$.
(a) attractive, and (b) repulsive lattice beams.
The horizontal axis is the local lattice depth $V_{0L}^{diag}$ which depends monotonically on the distance $r$ from the trap center along a $\{\pm 1,\pm 1, \pm 1\}$ direction.
The zero of energy for the vertical axis corresponds to the potential in
region $R_{III}$, where all the lattice and compensation lasers are negligible. The red stars denote the bottom of the band along a diagonal, and correspond to the lowest-energy single-atom
Bloch state when the system is empty.  The overall chemical potential $\mu$ is denoted by the horizontal black dashed lines.
If $\mu< 0$, $R_{III}$ is empty;
$\mu$ below the red stars corresponds to vacuum. The shaded blue regions correspond to the AF phase. $\mu$ outside of the blue region and above the red stars corresponds to a paramagnetic Fermi gas in reservoir $R_I$.
The green plus signs denote the bottom of the band in reservoir $R_{II}$, where only one lattice beam is present.  $R_{II}$ is empty when $\mu$ is below these.
Since the laser potentials in the different directions are additive, the band bottom along the $x$, $y$ or $z$ axis in $R_{II}$ (green plus signs) is one third of the band bottom along the diagonals (red stars).
The chemical potential is chosen so that the AF phase appears around the optimal lattice depth ($V_0=4 E_R$ for $a_s=0.1d$) calculated in the Hartree approximation \cite{Mathy2009}.
The AF phase occupies a relatively narrow region of values of lattice depths, which translates into a narrow region of the trap. As the lattice gets deeper, the bottom of the band gets pulled down (up) for attractive (repulsive) lattice beams.
}
\label{fig:musnocomp}
\end{figure}

\subsection{The model}
We consider two-component ultracold fermions of mass $m$ in a simple-cubic optical lattice interacting repulsively via a Feshbach resonance,
but far enough from the Feshbach resonance to apply the first order Born approximation. We call $d$ the lattice spacing, $a_s$ the $s$-wave scattering length,
and measure energy in units of the recoil energy $E_R= \hbar^2 k_R^2/(2m) = \hbar^2\pi^2/(2md^2)$ where $k_R=\pi/d$ is the recoil momentum. 
The Hamiltonian is
\begin{eqnarray}
H=\sum_{\sigma} \int d\vect{r} \hat{\Psi}^{\dagger}_{\sigma}(\vect{r}) \Big(-\frac{\hbar^2\nabla^2}{2m}&+&V(\vect{r}) \Big) \hat{\Psi}_{\sigma}(\vect{r}) \nonumber \\
&+& g\int d\vect{r} \hat{\rho}_{\uparrow}(\vect{r}) \hat{\rho}_{\downarrow}(\vect{r})
\end{eqnarray}
where $\vect{r}=\{x,y,z\}$; $\hat{\Psi}_{\sigma}(\vect{r})$ ($\hat{\Psi}^{\dagger}_{\sigma}(\vect{r})$) is the fermionic annihilation (creation) operator of spin $\sigma$ at position $\vect{r}$; $\hat{\rho}_{\sigma}(\vect{r})=\hat{\Psi}^{\dagger}_{\sigma}(\vect{r}) \hat{\Psi}_{\sigma}(\vect{r})$ are the density operators; and $V(\vect{r})$ is the total potential felt by the atoms. The first order Born approximation gives $g=4\pi \hbar^2 a_s/m$.


The total potential $V(\vect{r})$ is composed of an external potential
$V_{ext}(\vect{r})$, three lattice beams, and three compensating laser beams. The external potential $V_{ext}(\vect{r})$ (which may be zero) is generally provided by optical dipole forces and varies over a length scale much larger than
all the other length scales in the system. The three lattice beams and compensating beams are oriented along the $x$, $y$, and $z$ axes.
We call $V_{LC}(x;y,z)=V_{L}(x;y,z)+V_C(x;y,z)$ the sum of an optical lattice along the $x$ direction, produced by a retroreflected Gaussian laser beam, and a
non-retroreflected Gaussian beam along the $x$ direction, that serves to partially compensate the overall 
average potential of the lattice beams:
\begin{eqnarray}
V_{L}(x;y,z)&=&\mp V_{0L}\exp{(-\frac{2(y^2+z^2)}{w_{L}^2})}\sin^2{(k_R x)} \\
V_{C}(x;y,z)&=&\pm V_{0C}\exp{(-\frac{2(y^2+z^2)}{w_{C}^2})}~.
\end{eqnarray}
The intensities $V_{0L}$ and $V_{0C}$ of the lattice and compensating beams are positive.
The upper signs correspond to having attractive lattice beams and repulsive compensating beams, the lower signs to the opposite situation.
Generating a three-dimensional simple-cubic optical lattice requires three copies of the retroreflected lattice and compensating  beams, in the three orthogonal directions. The lattice beams alone lead to a simple-cubic optical lattice in a region of space that is limited by the Gaussian profile of the beams.
$w_{L}$ and $w_C$ are the waists of the lattice and compensating beams, respectively, and we define their ratio as $\alpha=w_{L}/w_C$.
The total potential is given by
\begin{eqnarray}
V(\vect{r})&=& V_{ext}(\vect{r}) + V_{LC}(x;y,z)\nonumber\\
&+& V_{LC}(y;z,x)+V_{LC}(z;x,y).
\end{eqnarray}

The position and size of the different phases in this trap are determined by combining a calculation of the phase diagram of a homogeneous system with the local density approximation (LDA).
In previous work \cite{Mathy2009}, two of us studied the phase diagram of this model in the Hartree approximation at zero temperature 
with a potential 
$V_{0L}\Big(sin^2(k_Rx)+sin^2(k_Ry)+sin^2(k_Rz) \Big)$. To relate it to the present work, we assume that $k_Rw_C\gg 1$ and $k_Rw_{L}\gg 1$, so that the Gaussian envelopes of the potentials vary much more slowly than the lattice spacing.
Under the LDA, the potential at each point in the trap is a sum of sinusoidal potentials plus an overall shift $\mu_{con}(x,y,z)$ 
that is the local minimum of the lattice potential:
\begin{eqnarray}
V(\vect{r})=\mu_{con}(x,y,z)+ V_{0x}(y,z) sin^2(k_Rx+\phi_x) &&\nonumber \\
 +V_{0y}(x,z) sin^2(k_Ry+\phi_y)+ V_{0z}(x,y)sin^2(k_Rz+\phi_z)&&
\end{eqnarray}
where $\phi_x, \phi_y, \phi_z$ are relative phases which are unimportant in the LDA. 
We neglect the spatial variations of the lattice amplitudes and $\mu_{con}$ over the period of the lattice oscillation, but we do account for their variation over the longer scales given by the beam waists $w_{L}$ and $w_C$. We choose the magnitudes of the sinusoidal parts of the potential to be positive: $V_{0x}(y,z)$, $V_{0y}(z,x)$, $V_{0z}(x,y) > 0$.
This can be done whether the lattice beams are repulsive or attractive by an appropriate choice of $\mu_{con}(x,y,z)$, $\phi_x$, $\phi_y$ and $\phi_z$, since for example
$-V_{0x}(y,z)sin^2(k_R x)=-V_{0x}(y,z)+(V_{0x}(y,z)) sin^2(k_R x-\pi/2)$, and we can absorb $-V_{0x}(y,z)$ into $\mu_{con}(x,y,z)$ and $-\pi/2$ into $\phi_x$.

The density in the trap is set by choosing an overall chemical potential $\mu$ for the system, assuming the system is at global equilibrium at zero temperature.
Within the LDA, the system sees a potential 
$V_{0x}(y,z) sin^2(k_Rx)+ V_{0y}(x,z) sin^2(k_Ry)+ V_{0z}(x,y)sin^2(k_Rz)$,
and has a local chemical potential given by $\mu-\mu_{con}(x,y,z)$.  
At each point in the trap the Hartree calculation takes the local chemical potential and lattice intensity and returns local 
properties such as density, staggered magnetization and the Mott-Hubbard gap. The Hartree calculations were restricted to the case of equal
lattice intensities, $V_{0x}=V_{0y}=V_{0z}$, 
which occurs along 4 straight lines in the $\{1,\pm 1,\pm 1 \}$ spatial directions.  
We call $V_{0L}^{diag}(r)$ the intensities of the lattice beams, and $\mu_{con}^{diag}$ the chemical potential shift due to the lasers along the $\{\pm 1,\pm 1, \pm 1\}$ directions, where $r=\sqrt{x^2+y^2+z^2}$. They have the following form:
\begin{eqnarray}
&&V_{0L}^{diag}(r)= V_{0L} exp(-\frac{4 r^2}{3w_{L}^2}) \label{eq:Vdiag}\\
&&\textrm{Attractive lattice beams}:\\
&&\mu_{con}^{diag}(r)=3\big(V_{0C} exp(-\frac{4 r^2}{3w_{C}^2})-V_{0L} exp(-\frac{4 r^2}{3w_{L}^2})\big) \\
&&\textrm{Repulsive lattice beams}: \\
&&\mu_{con}^{diag}(r)= 3\Big(-V_{0C} exp(-\frac{4 r^2}{3w_{C}^2})\Big)
\end{eqnarray}
We choose the zero of energy so $V_{ext}=0$ in the lattice (indeed we assume $V_{ext}$ is negligible in regions where other laser potentials are sizeable). 
We define a dimensionless parameter $\beta$ to characterize the ratio of intensities of the lattice and compensating beams: $(V_{0C}/E_R) exp(-\frac{4 r^2}{3w_{C}^2})=\beta \Big((V_{0L}/E_R)exp(-\frac{4 r^2}{3w_{L}^2}) \Big)^{\alpha^2}$, so that
\begin{equation}
V_{0C}/E_R=\beta (V_{0L}/E_R)^{\alpha^2}.
\label{eq:powerlaw}
\end{equation}
Thus $\beta$ gives the ratio of the intensities of the compensating and lattice beams at the point where $V_{0L}^{diag}(r)=E_R$. The potential on the $z=0$ surface is depicted schematically in Fig. \ref{fig:LaserCut}.

\subsection{The N\'{e}el state and its reservoirs}

The Hartree calculation in the lattice gives the regions of parameter space where the ground state is the N\'{e}el antiferromagnetic phase (AF), and
the highest-energy occupied and lowest-energy unoccupied Hartree single-atom states provides an estimate of the Mott-Hubbard gap (neglecting spin-wave corrections).
If the chemical potential lies anywhere within this gap, the phase will be AF.  Surrounding this AF phase is a ``reservoir" of atoms in a paramagnetic Fermi gas.
We want the atoms in this reservoir to be mobile so that they can carry away entropy from the part of the trap containing the AF phase.  Thus, we want the optical lattice to remain relatively weak in the reservoir.

We can distinguish between three different types of reservoir (Fig.~\ref{fig:LaserPhases}): $R_I$ is in the three-dimensional part of the lattice,
i.e. in the region where all lattice and compensating beams are non-negligible; $R_{II}$ are the 6 reservoirs in the regions where the beam intensities are appreciable
in only one direction, corresponding to taking one of coordinates $|x|$, $|y|$ or $|z|$ 
large compared to the beam waists while leaving the other two small enough to remain within the beam; and $R_{III}$ is the reservoir outside of all of the beams.



The parameters in the potential allow for significant freedom in tailoring the distribution of phases in the trap.
As one goes along one of the diagonal directions away from the origin, the amplitude of the lattice decays according to Eq. \ref{eq:Vdiag},
which sets the local effective amplitude $V_{0L}^{diag}$ of the simple-cubic optical lattice. There is no need to specify the waist of the lattice beam within the LDA,
as this simply sets the linear size of the different phases in the trap.
The parameters that must be chosen are the ratio of beam waists $\alpha$ and the ratio of the intensities of the lattice and compensating beams $\beta$, as defined in Eq. \ref{eq:powerlaw}.

The chemical potential $\mu$ can be directly related to the density in the region $R_{III}$ where all laser potentials, except possibly for $V_{ext}(\vect{r})$,
are zero (see Fig.~\ref{fig:LaserPhases}).  If $\mu\leq V_{ext}(\vect{r})$ then $R_{III}$ is empty. 
For $\mu\geq V_{ext}(\vect{r})$ the Hartree approximation gives the local density in $R_{III}$:
\begin{equation}
(\mu-V_{ext}(\vect{r}))/E_R= \frac{1}{\pi^2} (3\pi^2 nd^3)^{2/3}+\frac{8}{\pi} \frac{a_s}{d} nd^3~.
\label{eq:mu}
\end{equation}




\subsection{Optimizing the parameters}

The parameters of the system should be chosen to maximize the size of the phase of interest, and create optimal conditions for the realization of the phase.
The lattice depth which maximizes the Hartree estimate of the effective AF exchange $J_H$, thus giving the fastest equilibration timescales and maximum N\'{e}el temperature, for a given interaction strength $a_s$, was estimated in previous work \cite{Mathy2009}, and is plotted in Fig. \ref{fig:JH} as a function of lattice depth for the interaction strengths considered in this work: $a_s=0.1d$ and $a_s=0.06d$. We also found that the lattice depth which maximizes the entropy of the N\'{e}el state, in a calculation that neglects terms in the Hamiltonian beyond the Hubbard model \cite{Werner05}, is close to this optimal AF exchange lattice depth. Therefore, the center of the trap should be at a lattice depth close to this optimal lattice depth.
The AF phase should also occupy as large a volume as possible in the trap.



Figure \ref{fig:Potentials} schematically depicts the phases encountered in going along a diagonal from $r=0$, where the lattice is deepest,
to the edge of the trap, where the lattice depth goes to zero.  Figures \ref{fig:Potentials}(a) and (b)  show the cases of lattice potentials with no compensation, 
while Figs.~\ref{fig:Potentials}(c) and (d) show two cases of compensated lattice potentials. 
Without compensation the chemical potential within the AF phase depends strongly on lattice depth, and consequently, the AF phase occupies a narrower region of the trap.  One desired effect of compensation is to flatten the chemical potential of the AF phase with varying lattice depth, in order to enlarge the AF region.  In addition, as shown in Figs.~\ref{fig:Potentials}(c) and (d), compensation allows adjustment of $\mu$ with respect to $E=0$ (the potential outside of the lattice).  For $\mu$ slightly less than 0, $R_{III}$ is empty, and atoms may evaporate from the edges of the lattice, while for $\mu$ slightly above 0, $R_{III}$ is occupied and forms a thermal reservoir that may itself be evaporatively cooled and equilibrate with atoms at the lattice edge.

\begin{figure}[t!]
\centering
\parbox{1\linewidth} {
\subfigure{ \includegraphics[width=1\linewidth, clip=true,trim= 0 0 0 00]{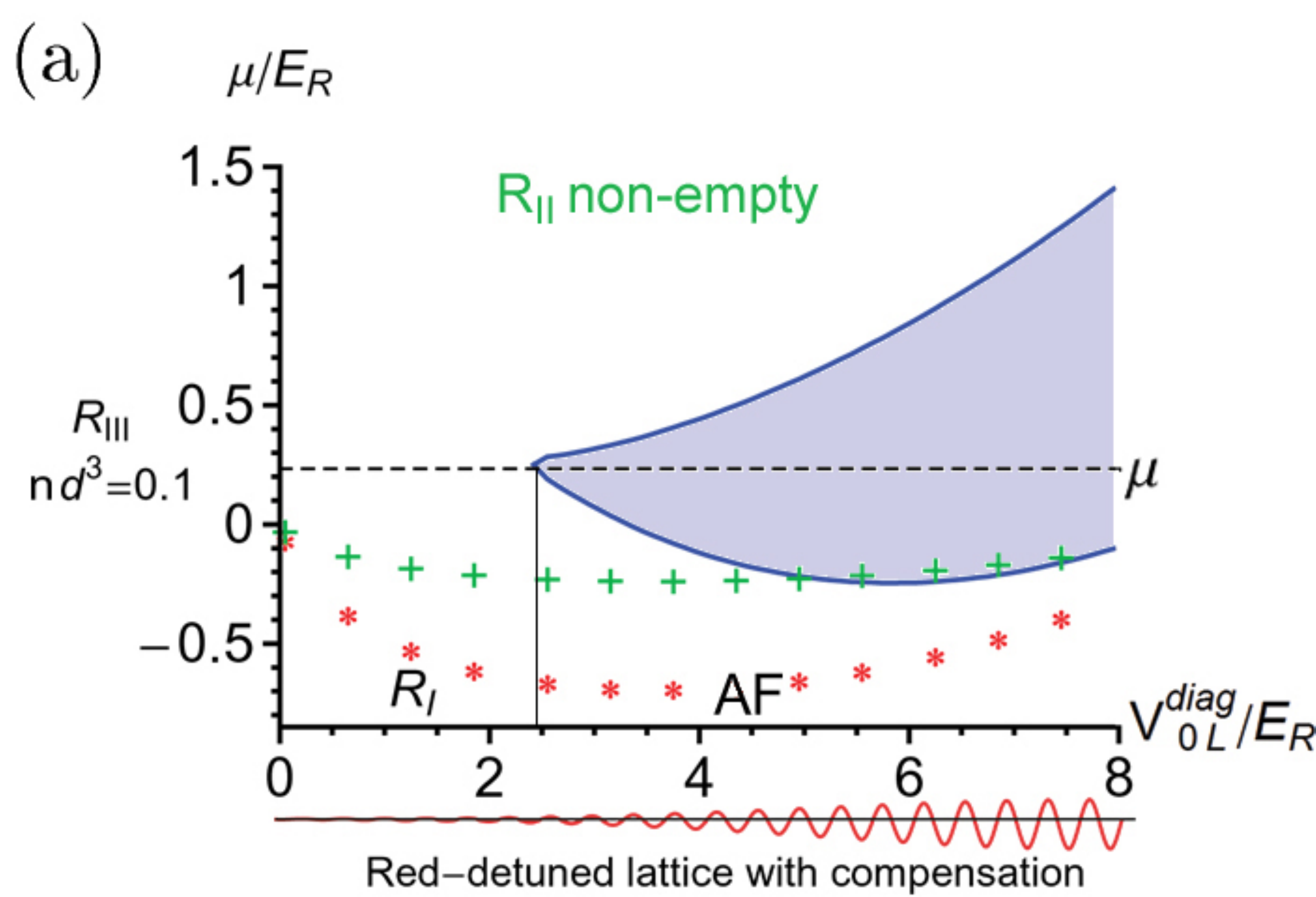}
\label{fig:musredcompRIII}
}
}
\qquad
\parbox{1\linewidth}
{
\subfigure{
\includegraphics[width=1\linewidth, clip=true,trim= 0 0 0 0]{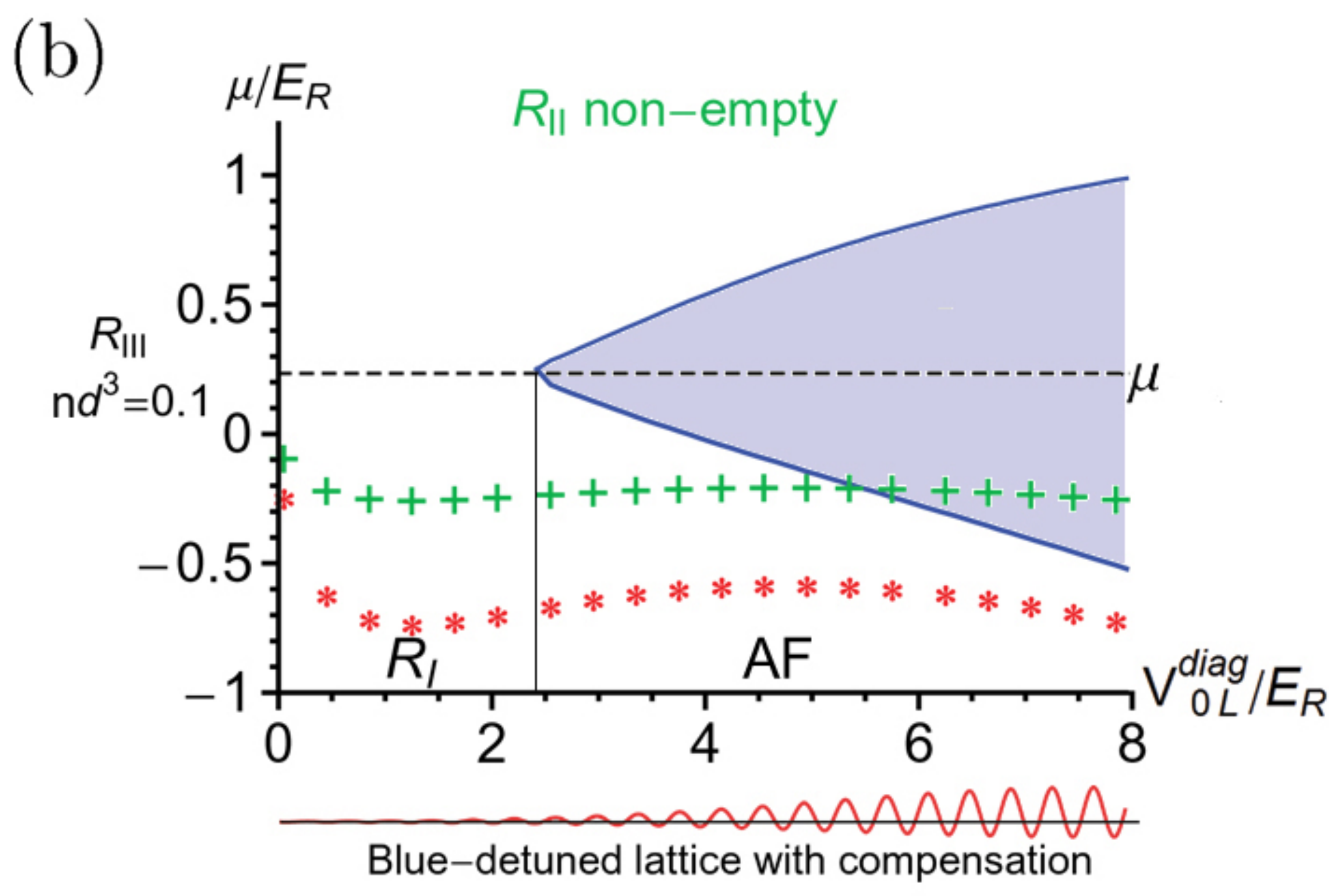}
\label{fig:musbluecompRIII}
}
}
\caption{
Phases for $a_s=0.1d$ with compensating beams, for (a) attractive, and (b) repulsive lattice beams, with $\mu >0$ chosen to give density is $nd^3=0.1$ in $R_{III}$ . The parameters are
(a) $\beta=0.379$, $\alpha=1.13$; and (b) $\beta=0.705$, $\alpha=0.81$.
The lines and symbols are defined as in Fig.~\ref{fig:musnocomp}.
}
\label{fig:muscompRIII}
\end{figure}

\begin{figure}[t!]
\centering
\parbox{1\linewidth} {
\subfigure{ \includegraphics[width=1\linewidth, clip=true,trim= 0 0 0 00]{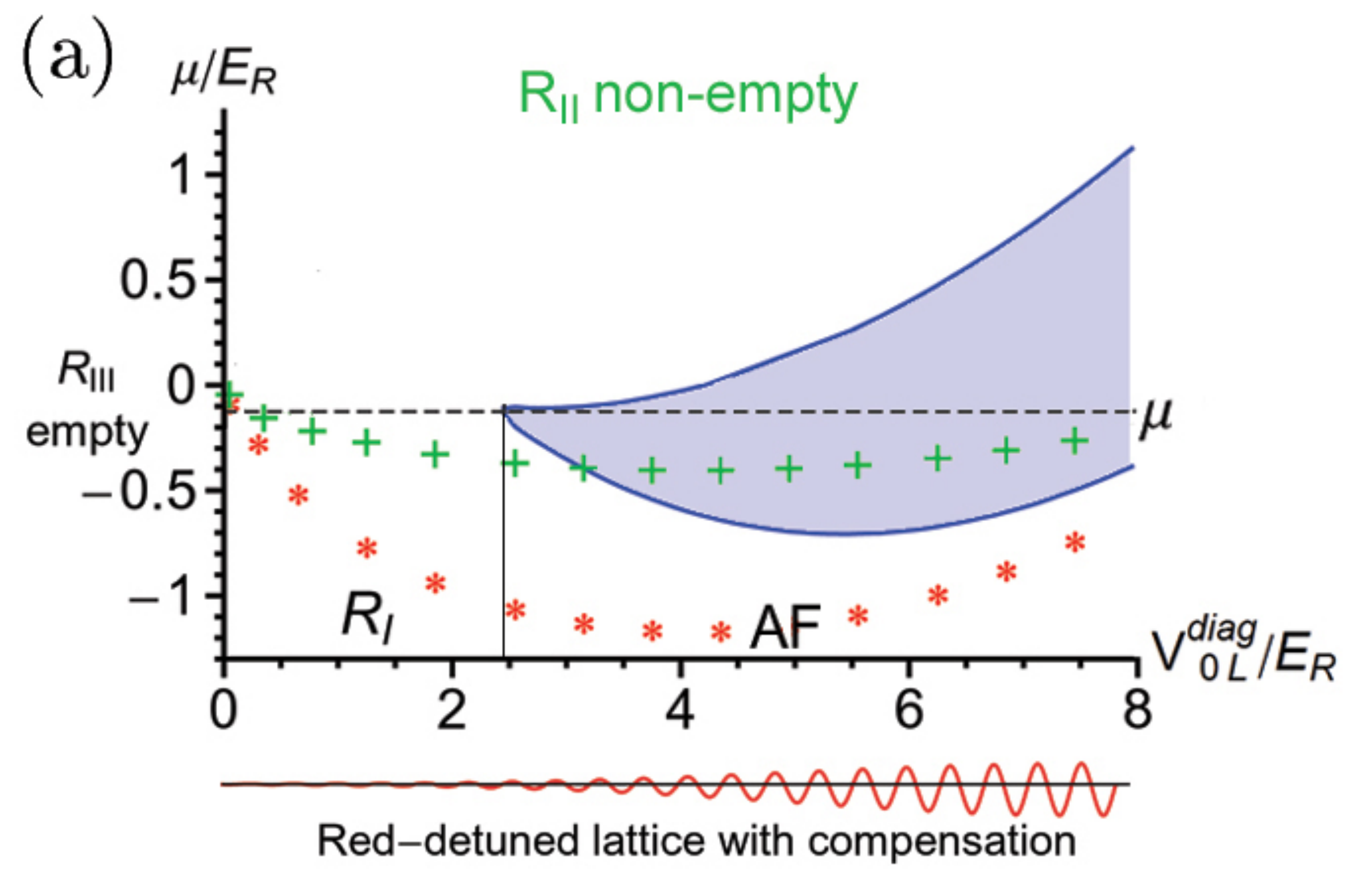}
\label{fig:musredcompNoRIII}
}
}
\qquad
\parbox{1\linewidth}
{
\subfigure{
\includegraphics[width=1\linewidth, clip=true,trim= 0 0 0 0]{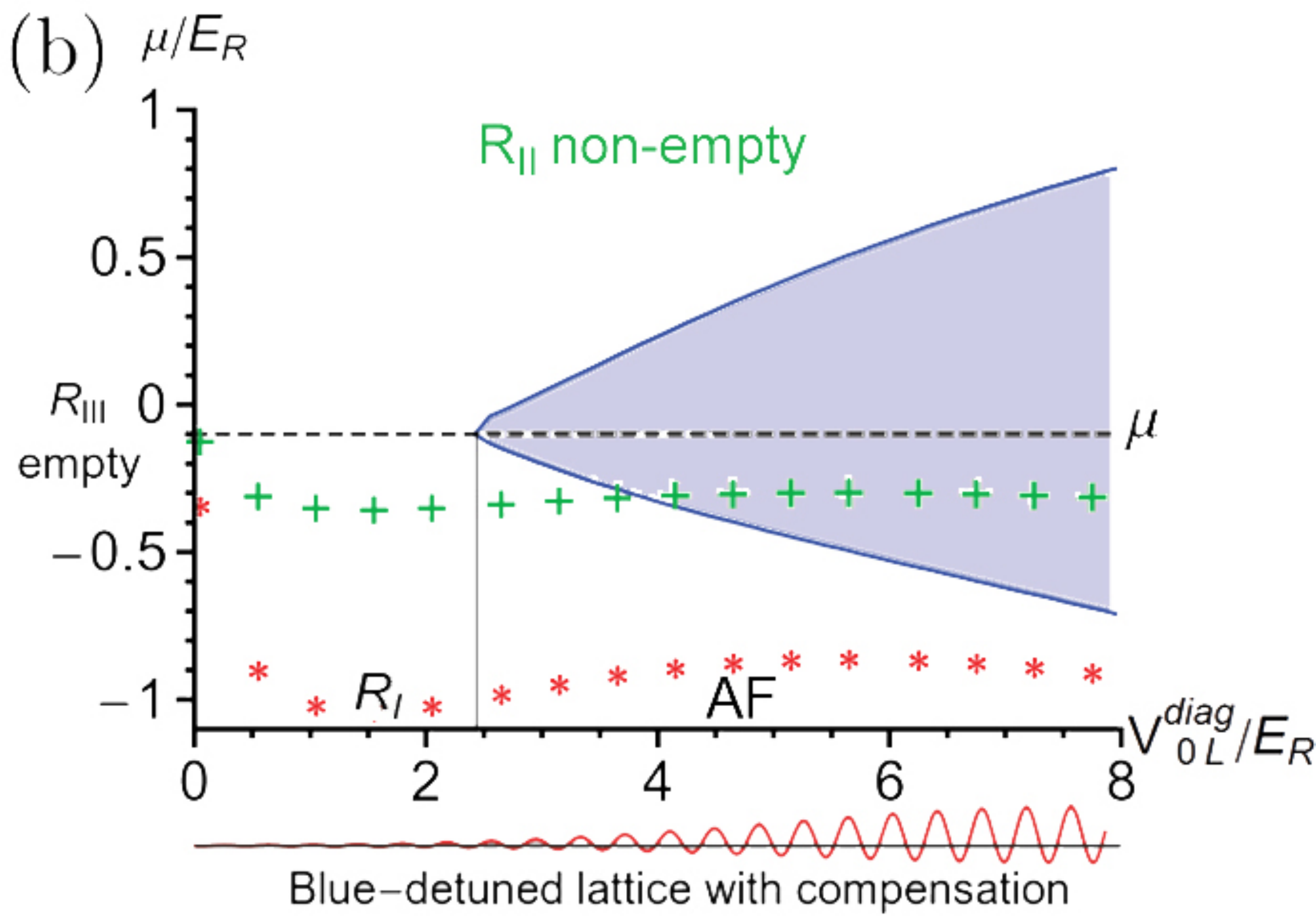}
\label{fig:musbluecompNoRIII}
}
}

\caption{(Color online)
Empty $R_{III}$ reservoir with $a_s=0.1d$.
(a) Attractive lattice with $\beta=0.315$, $\alpha=1.17$; (b) Repulsive lattice with $\beta=0.80$, $\alpha=0.77$.
}
\label{fig:muscompNoRIII}
\end{figure}

\begin{figure}[t!]
\centering
\parbox{1\linewidth} {
\subfigure{ \includegraphics[width=1\linewidth, clip=true,trim= 0 0 0 00]{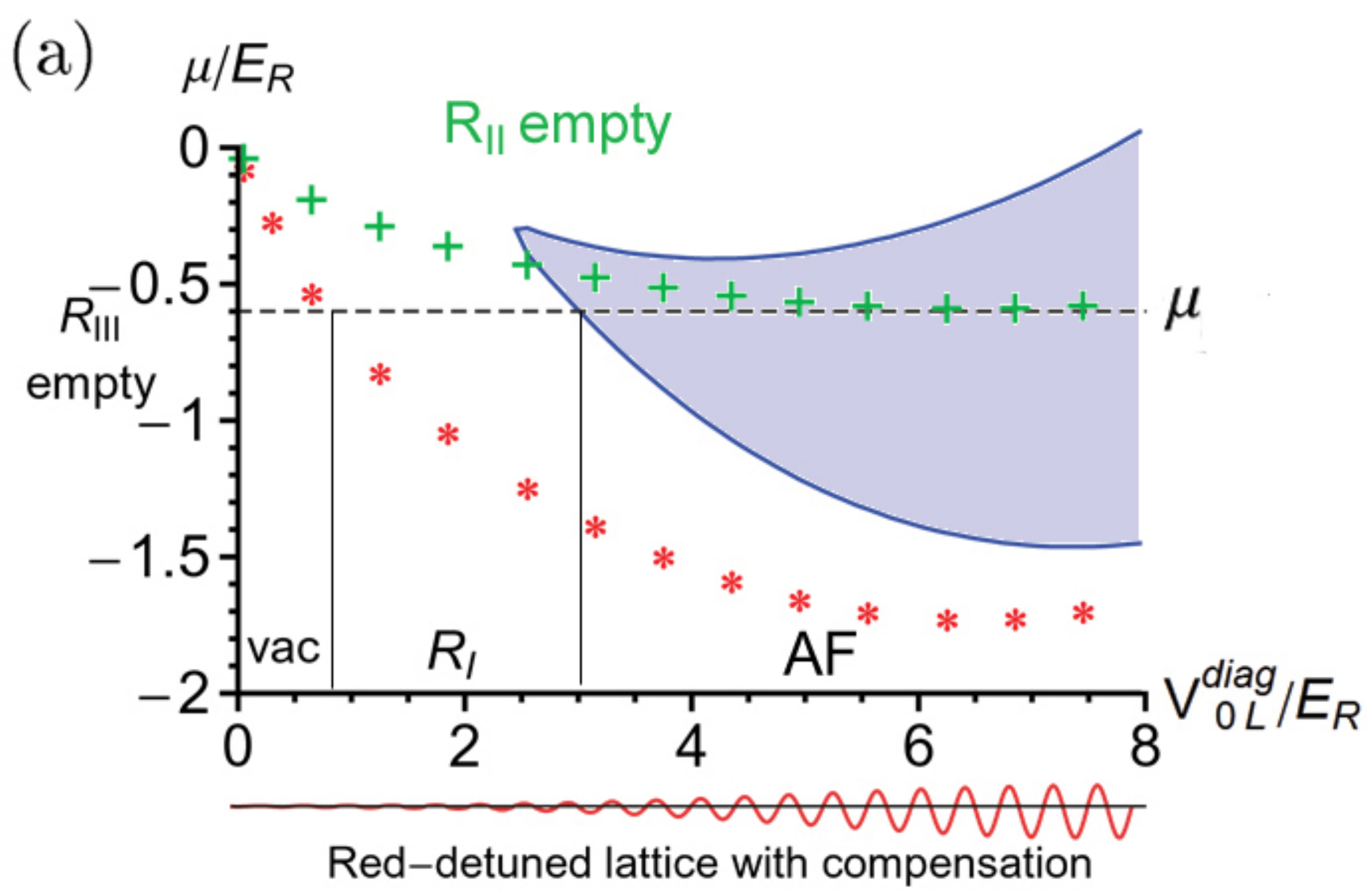}
\label{fig:musRIIEmpty}
}
}
\qquad
\parbox{1\linewidth}
{
\subfigure{
\includegraphics[width=1\linewidth, clip=true,trim= 0 0 0 0]{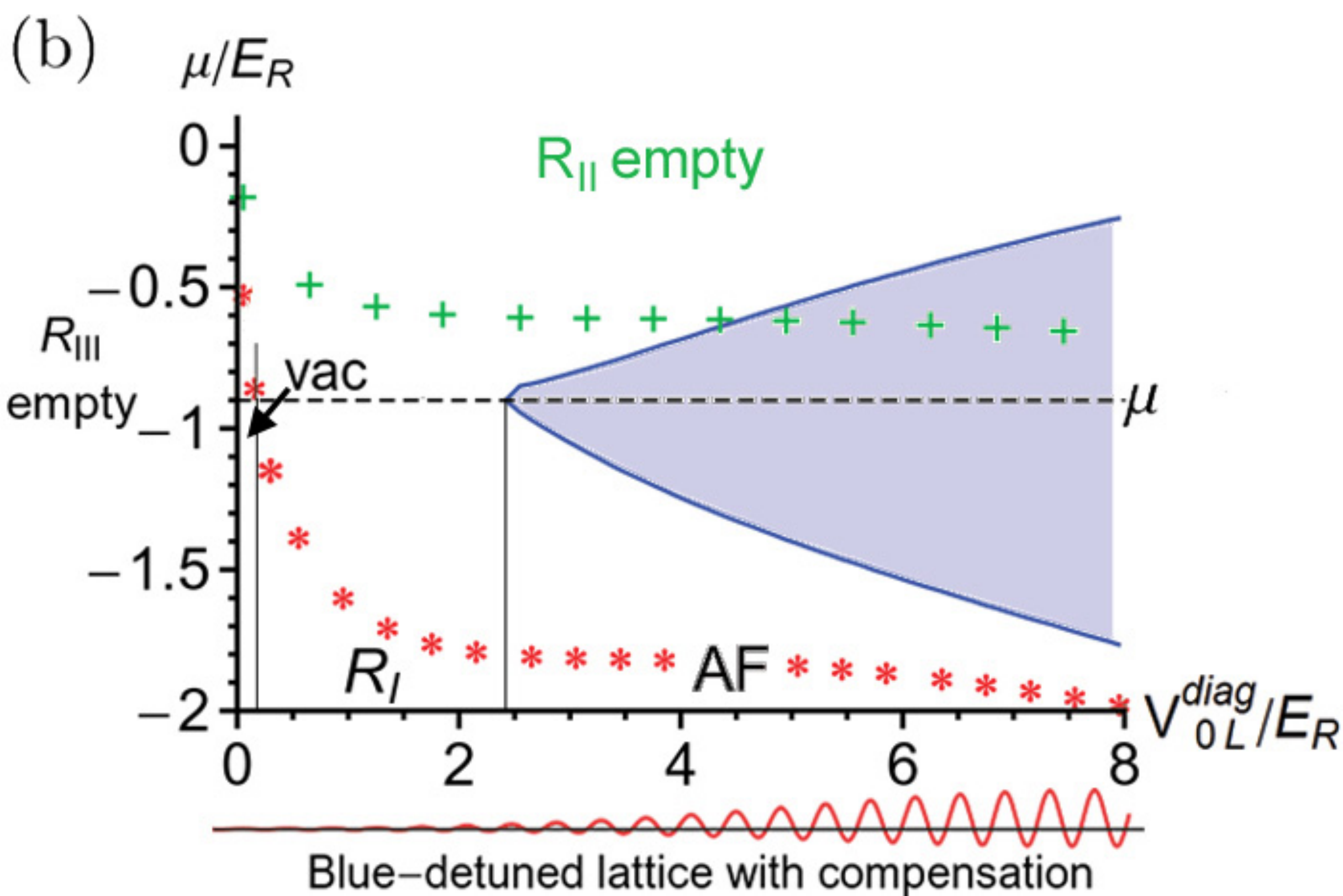}
\label{fig:musFlippedRIIEmpty}
}
}
\caption{(Color online)
Both $R_{II}$ and $R_{III}$ are empty, with $a_s=0.1d$.
(a) Attractive lattice with $\beta=0.30$, $\alpha=1.16$.  In this case, the AF phase does not go all the way to $2.4 E_R$.
(b) Repulsive lattice with $\beta=1.0$, $\alpha=0.74$.  For this setup in both (a) and (b), the evaporating atoms go out the ``beams'' through the
empty $R_{II}$'s and not through $R_{III}$, which is at too high an energy relative to $\mu$.
}
\label{fig:diffwaistsRIIEmpty}
\end{figure}

\begin{figure}[t!]
\centering
\parbox{1\linewidth} {
\subfigure{ \includegraphics[width=1\linewidth, clip=true,trim= 0 0 0 00]{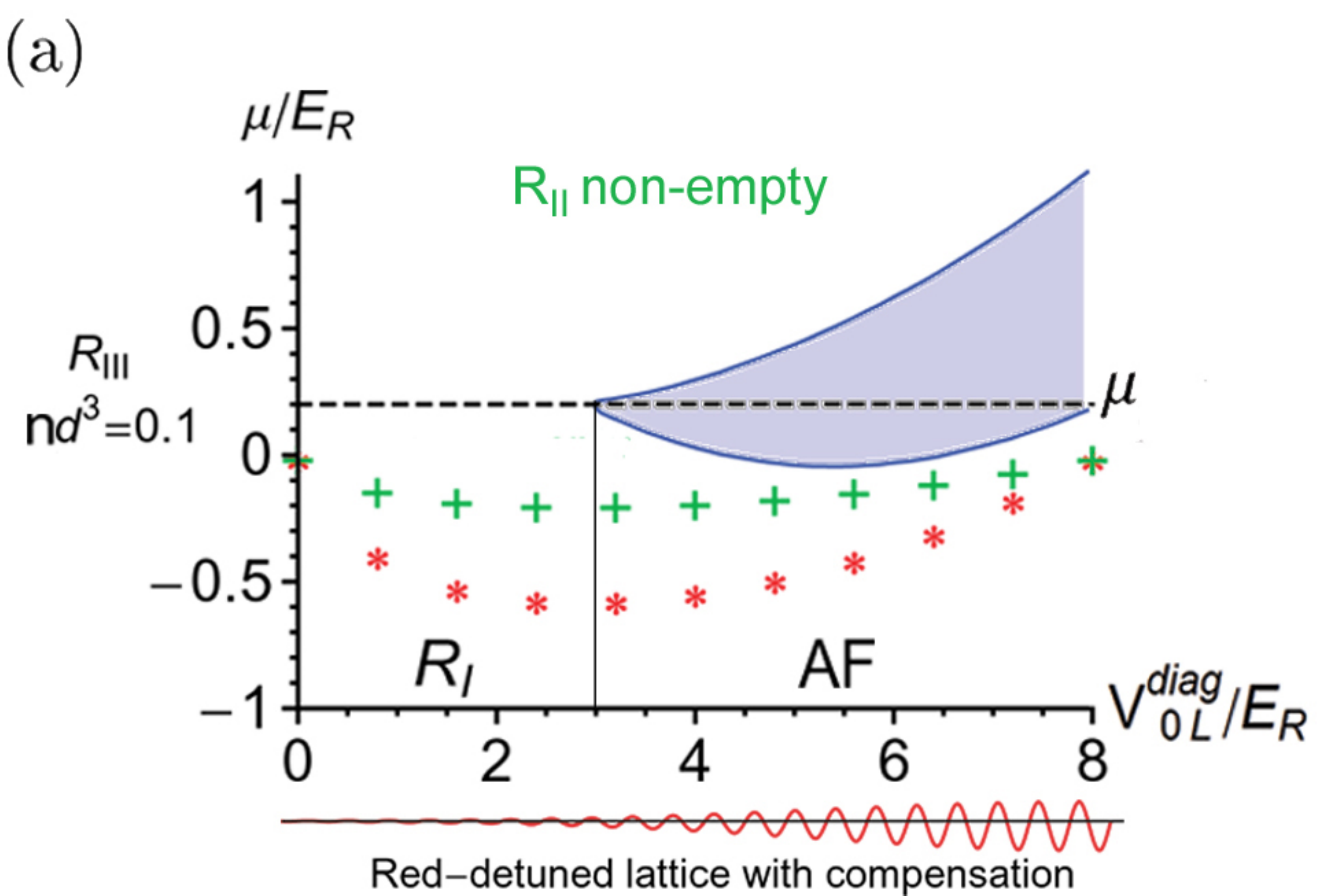}
\label{fig:musas0p06RIIIFull}
}
}
\qquad
\parbox{1\linewidth}
{
\subfigure{
\includegraphics[width=1\linewidth, clip=true,trim= 0 0 0 0]{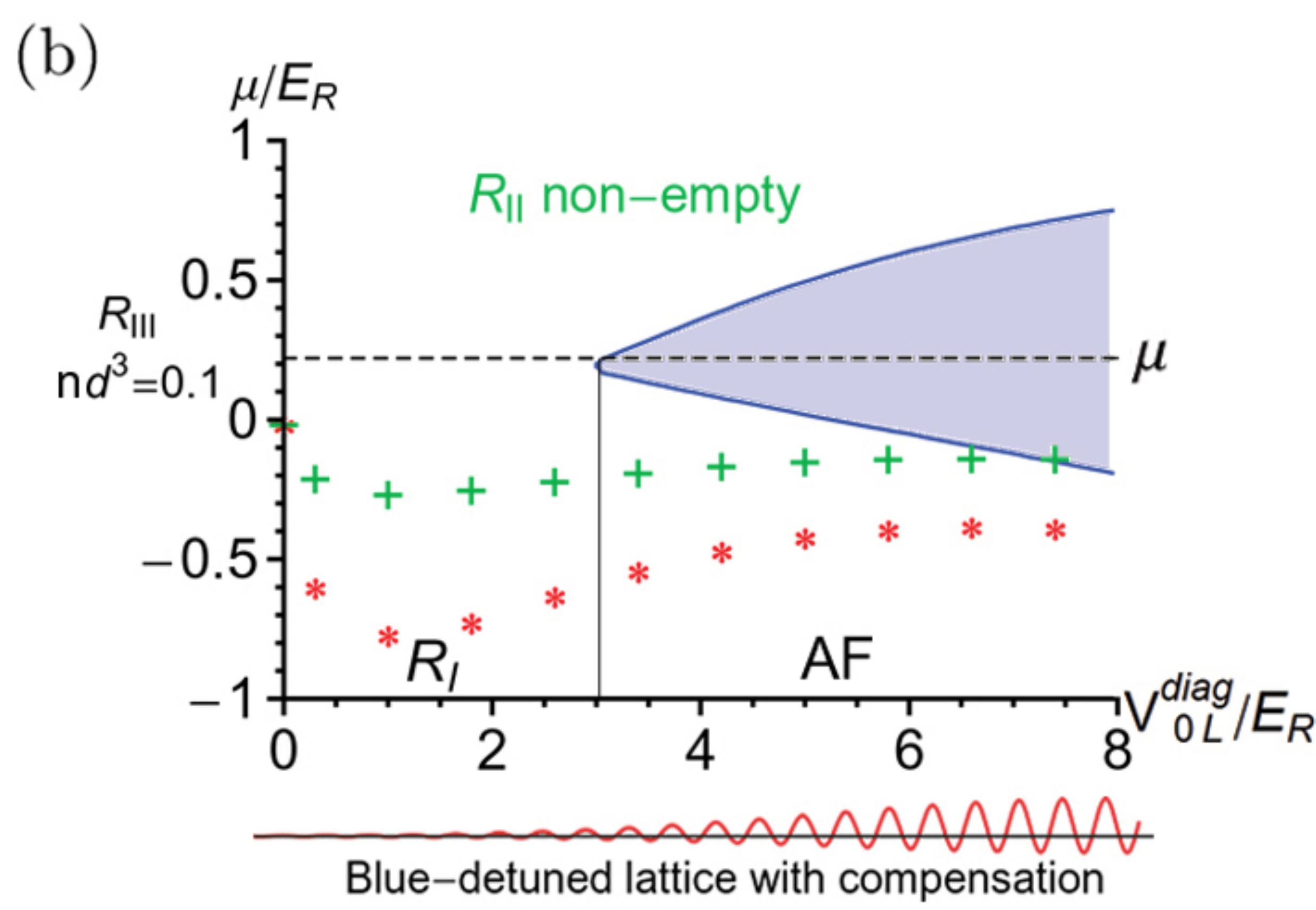}
\label{fig:musFlippedas0p06RIIIFull}
}
}

\caption{(Color online) Similar situation to Fig.~\ref{fig:muscompRIII}, with a lower scattering length:
$a_s=0.06 d$.
(a) Attractive lattice with $\alpha=1.13$, $\beta=0.389$;
(b) Repulsive lattice with $\alpha=0.79$, $\beta=0.72$.
}
\label{fig:musa0p06}
\end{figure}

  More quantitatively, the results of our calculation of the phases are shown in
Figures \ref{fig:musnocomp} to \ref{fig:musa0p06}.
In these plots, the horizontal axis is the lattice depth along a diagonal from the edge of
the trap to the center.  The center of the trap can be chosen to be anywhere along this axis (or beyond).  Figure \ref{fig:musnocomp} shows the cases with no compensating beams.  The smallness of the AF region due to the strong variation of the
AF phase with lattice depth is readily apparent.
To make the AF phase occupy a larger fraction of the trap, the compensating beam must
shift the chemical potential of the AF phase so that it remains constant with varying lattice depth as shown in the following figures.

Three different scenarios can be engineered for the fillings of the reservoirs, which we consider for $a_s=0.1d$ in the following plots: $R_{III}$ non-empty (Fig.~\ref{fig:muscompRIII});
$R_{III}$ empty (Fig.~\ref{fig:muscompNoRIII}); and $R_{II}$ and $R_{III}$ empty (Fig.~\ref{fig:diffwaistsRIIEmpty}).
To get a sense of the dependence on interaction strength, we plot the phases for $a_s=0.06 d$ when all three reservoirs are occupied in Fig.~\ref{fig:musa0p06}. 

\begin{figure}[t!]
\centering
\parbox{1\linewidth} {
\subfigure{ \includegraphics[width=1\linewidth, clip=true,trim= 0 0 0 00]{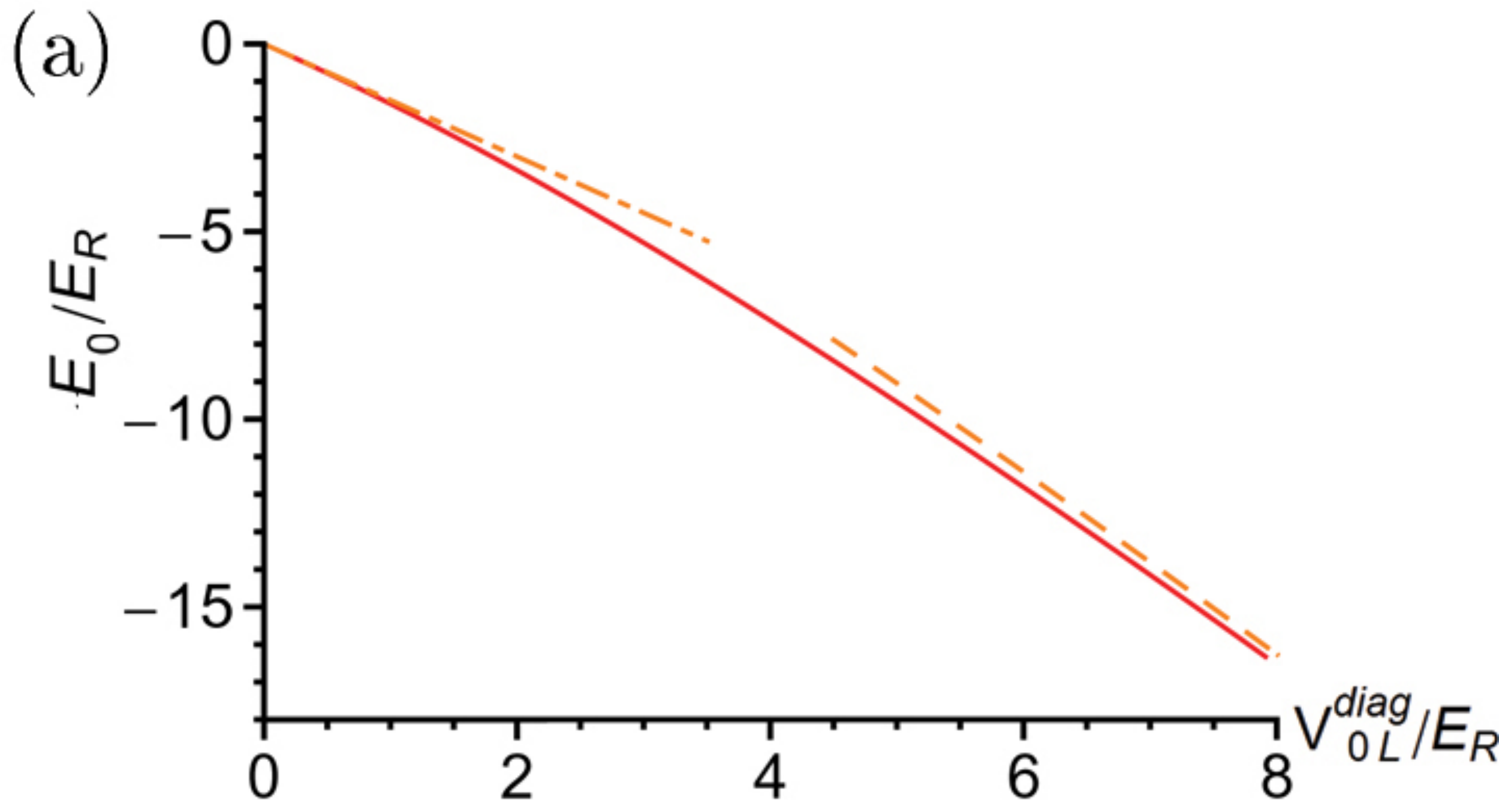}
\label{fig:BandBottomAttractive}
}
}
\qquad
\vspace{0.2in}
\parbox{1\linewidth}
{
\subfigure{
\includegraphics[width=1\linewidth, clip=true,trim= 0 0 0 0]{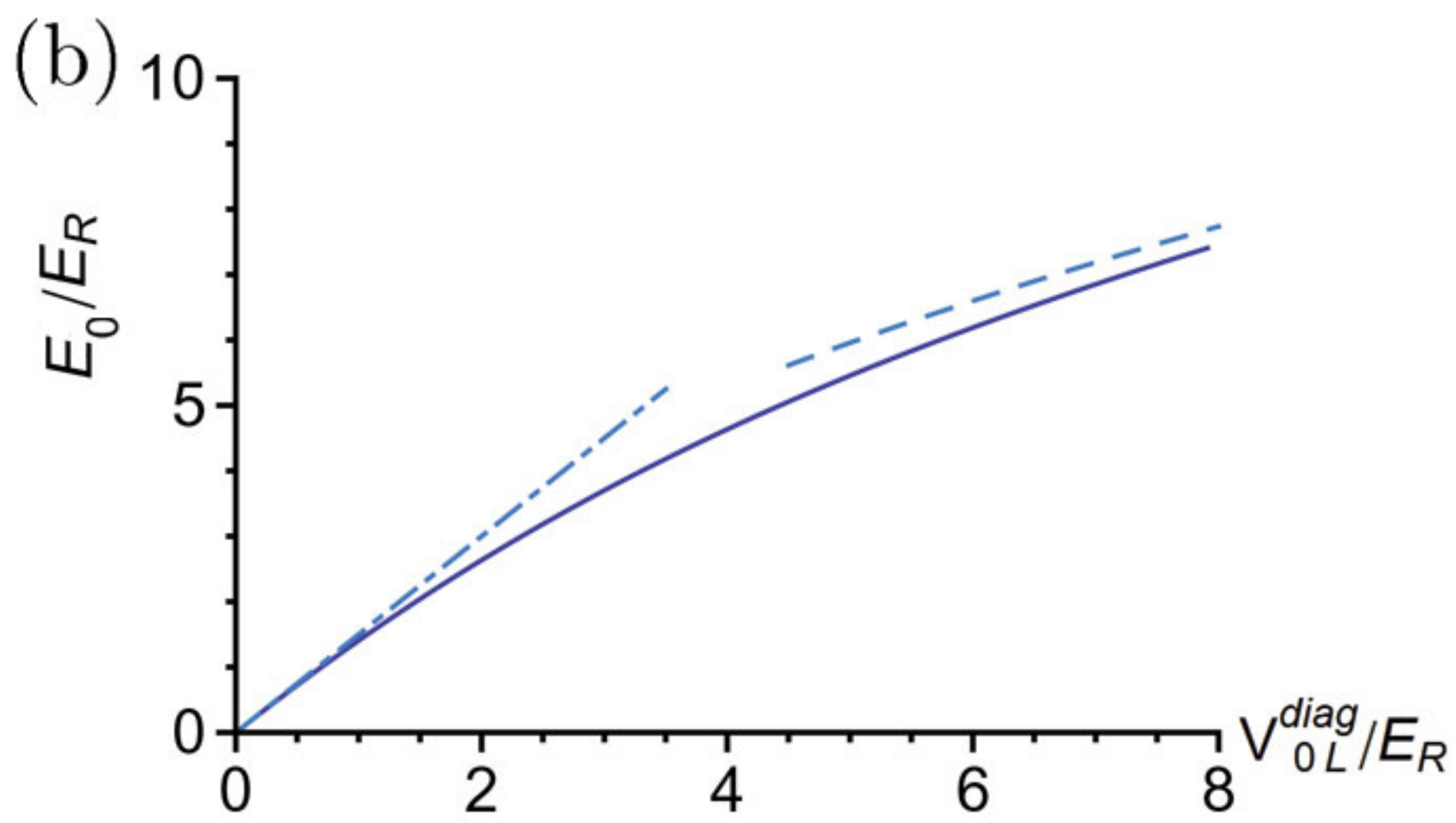}
\label{fig:BandBottomRepulsive}
}
}
\caption{(Color online)
 Bottom of the lowest Bloch band along the diagonal directions, i.e. for a potential $V(x,y,z)=V_{0L}^{diag} ( sin^2(k_R x)+sin^2(k_R y)+sin^2(k_R z))$ for (a) attractive lattice beams ($V_{0L}^{diag}<0$) and (b) repulsive lattice beams ($V_{0L}^{diag}>0$). The energy difference between the bottom of the lowest Bloch band and the bottom of the Mott gap decays quickly with increasing lattice depth. Therefore, to keep the chemical potential in the Mott gap one has to flatten  the bottom of the band, for deep lattices.
(a) Full red line: bottom of the band in the case of attractive lattice beams. The dashed (dot-dashed) line gives the asymptotic behavior at deep (weak) lattice. For a weak lattice, first-order perturbation theory gives the bottom of the gap to be $-3 V_{0L}^{diag}/2$. For a deep lattice, using the harmonic and lowest order anharmonic terms for the wells in the lattice, one gets that the band bottom goes like $-3V_{0L}^{diag}+3 (\sqrt{V_{0L}^{diag}/E_R}-1/4) E_R$. The band bottom goes down superlinearly in lattice depth, so the compensating beam intensity must therefore grow superlinearly in $V_{0L}^{diag}$.
(b) Full blue line: bottom of the band in the case of repulsive lattice beams. The dashed (dot-dashed) line gives the asymptotic behavior at deep (weak) lattice. For a weak lattice, the band bottom goes as $\sim 3 V_{0L}^{diag}/2$, while for a deep lattice, it becomes $\sim 3 (\sqrt{V_{0L}^{diag}/E_R}-1/4) E_R$. Therefore, the band bottom grows sublinearly in lattice depth, so the compensating beam intensity must grow sublinearly in $V_{0L}^{diag}$.
}
\label{fig:BandBottoms}
\end{figure}

A qualitative understanding of why different waists for the lattice beams and the compensating beams is advantageous is illustrated by Fig.~\ref{fig:BandBottoms}. Unequal beam waists will cause the strength of the compensating beams to grow as a power law in the strength of the lattice beams, as expressed by Eq.~\ref{eq:powerlaw}.
For deep lattices, the bottom of the Mott gap becomes narrowly separated from the bottom of the lowest Bloch band.  The objective of the compensating beams is to keep the chemical potential inside the Mott gap as the lattice depth varies.  In the absence of the compensating beams, the dependence of the bottom of the Mott gap on lattice depth is reasonably well described by a power law, so the compensating beams can flatten the Mott gap, thus achieving our objective.
Furthermore, we expect from the behavior of the band bottoms as a function of lattice depth shown in Fig.~\ref{fig:BandBottoms} that we need $\alpha>1$ for attractive, and $\alpha<1$ for repulsive lattice beams to obtain as flat a Mott gap as possible with the given setup.
The special case where the lowest order confinement is quartic  \cite{Ma08} is achieved by setting  $V_{0L}/V_{0C} = \alpha^2$. While this choice of parameters flattens the bottom of the potential, it does not maximize the volume of the N\'eel phase, since that is achieved by flattening the Mott gap instead, as we propose.


The Hartree approximation we have used in this work is a mean-field approximation and the real system will be quantitatively somewhat different from these Hartree estimates, due to nontrivial fluctuations and correlations.  However, we do not require precise numerical results to show the effectiveness of the presented scheme.  The Mott-Hubbard gaps are large enough in the region of interest that quantitative changes in the precise values will not destroy the general qualitative features that this scheme relies upon.  The solid (blue) lines in Figures \ref{fig:musnocomp} to \ref{fig:musa0p06}
bound the AF Mott insulating phase.  The Hartree approximation probably overestimates the range of stability of this phase, since it does not include all quantum fluctuations.  But if the AF phase is quantitatively a little smaller in these figures than what we show, the chemical potential and compensating potential can still be adjusted to enlarge the AF phase to fill the entire central region of the lattice, and allow continued evaporation from the AF phase within the lattice.

\subsection{Maximizing the Bragg signal}

\begin{figure}[t!]
\includegraphics[width=1\linewidth]{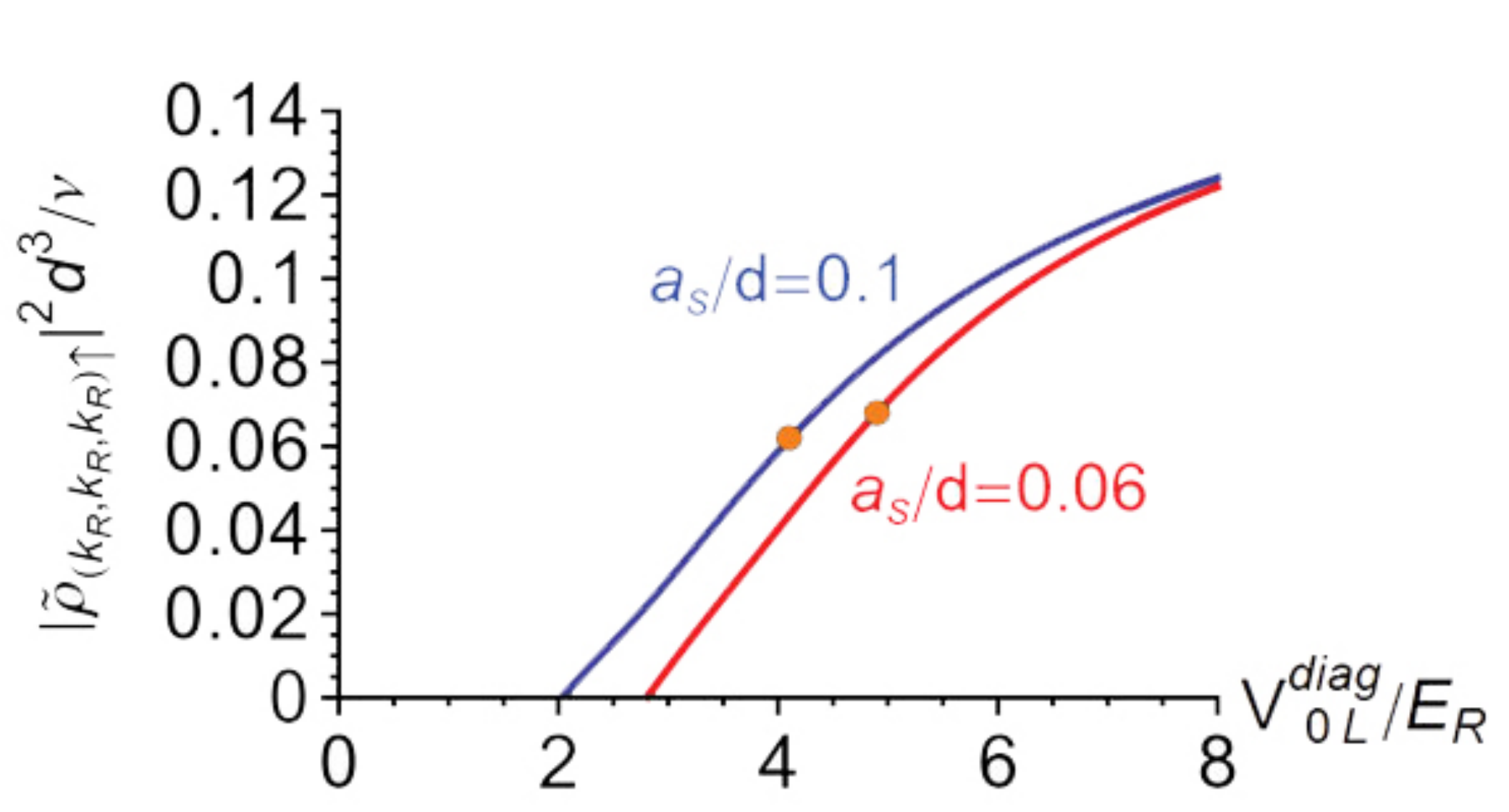}
\caption{ (Color online)
Square of the Fourier transform at momentum $(k_R,k_R,k_R)=(\pi/d,\pi/d,\pi/d)$ of the up spin density, where $d$ is the lattice spacing, and $\nu$ is the volume of the system, for two interaction strengths: $a_s=0.1d$ (blue) and $a_s=0.06d$ (red). The Bragg signal obtained from scattering light off the up spins will be proportional to this quantity. At infinite lattice depth, it becomes $1/4$, as the Fourier transform of infinitely localized particles on the FCC lattice at this momentum is 1/2. The highlighted (orange) points are at the corresponding lattice depth where the AF superexchange is maximized. While a deeper lattice and stronger interactions lead to more localized particles, and therefore a stronger Bragg signal, deeper lattices also lead to smaller superexchange and therefore smaller ordering temperatures and longer timescales for heat transport.
Both objectives can be met by cooling with a weaker lattice and then ramping up the lattice
before performing Bragg scattering.
}
\label{fig:Bragg}
\end{figure}

The AF phase can be directly detected using
Bragg scattering of near-resonant light \cite{Corcovilos10}. The Bragg signal from scattering off
the up spin density, for example, is proportional to the volume of the AF phase and to the square of the Fourier transform of the up spin density
at momentum  $(k_R,k_R,k_R)$. We have shown that the volume of the AF phase may be maximized by varying the relative intensities and waists of the lattice and
compensating beams. While cooling and equilibration times are minimized at relatively low lattice depths, the Bragg signal is enhanced at deeper lattices,
for which quantum fluctuations due to the site being doubly- or unoccupied are weaker. Figure \ref{fig:Bragg} shows a plot of the Fourier
intensity in the ground state as a function of lattice depth for $a_s=0.1d$ and $a_s=0.06d$. The lattice depths which maximize AF superexchange
in the Hartree approximation\cite{Mathy2009} are indicated. We see that the Bragg signal is maximized by going to deeper lattices and stronger interactions. Therefore,
one must compromise between the conditions that minimize the time scales for equilibration and cooling and those that maximize the Bragg signal.

One way to strengthen the Bragg signal is to cool and equilibrate at the relatively low lattice depths that maximize superexchange, but before performing Bragg scattering ramp up the lattice depth \cite{Koetsier08,Dare07} at a speed that is sufficiently adiabatic to reduce the quantum fluctuations. Since these fluctuations arise from virtual pairs of empty and doubly-occupied sites due to the superexchange process, one should be able to remove them provided that the lattice ramp is adiabatic with respect to the Mott-Hubbard gap. The spin-wave zero-point fluctuations that are present in the corresponding Heisenberg model are not strongly reduced in the limit of a deep lattice, while the virtual vacancies and doubly occupied sites are strongly suppressed. At nonzero temperature thermal fluctuations will also produce real empty and doubly-occupied sites. The lattice ramp will bias the hopping of these thermally-excited site defects and possibly produce heating unless the compensating beams are carefully ramped together with the lattice to eliminate such forces. The ramp should be fast enough to shut down the hopping in order to freeze in these thermal excitations before they can recombine and release their energy,  which becomes increasingly high compared to the energy of spin fluctuations as the lattice strengthens.  This suggests an optimal ramp rate is the fastest possible while remaining adiabatic with respect to the Mott-Hubbard gap in the bulk of the AF phase.
The ramp will be nonadiabatic near the outer edges of the AF phase, where the lattice is weak and the Mott gap is initially very small, causing some of the Bragg signal to be lost, but since the N\'eel ordering was initially very weak there the gain by enhancing the Bragg signal over the bulk of the AF phase should outweigh this loss near its edges.  The precise balance between these various nonequilibrium dynamical considerations is a challenge, and deserves further study.

Bragg scattering relies on the antiferromagnetic ordering being along the spin direction set by the ``up''
and ``down'' hyperfine states, which we call the $z$-direction. The local up and down spin populations may not be precisely equal, however, because of fluctuations in the initial conditions.
This local spin polarization along the $z$ direction
produces canted antiferromagnetism \cite{Wunsch10,Snoek11}, in which the AF order 
is tilted only slightly away from the $xy$ plane.  In this case, a $\pi/2$ pulse before the Bragg measurement will
tip the AF order up to a plane containing the $z$-direction, 
making it detectable with Bragg scattering \cite{Corcovilos10}.

If the lattice depth is ramped up before Bragg scattering, the $\pi/2$ pulse should be applied before the lattice ramp for the following reasons:
At deep lattices, the spin-spin interaction is greatly reduced and thus the components of the spins pointing in the $xy$ plane on different lattice sites may dephase 
with respect to one another due to thermal and quantum fluctuations.
However, spins pointing in the $z$-direction will not dephase since they are eigenstates of the single-atom Hamiltonian in a single deep well.  Therefore, a $\pi/2$ pulse before ramping the lattice depths will partially prevent dephasing of the antiferromagnetic correlations that can occur at large $V_{0L}$, by increasing the amount of the antiferromagnetic order that is along the $z$-direction.

\subsection{Conclusions}

We have proposed a setup to 
facilitate both realizing and detecting the N\'{e}el state of two-component fermions in a simple-cubic optical lattice. We found that the introduction of compensating
beams with a different beam waist allows for a significant growth of the N\'eel phase in the trap, and control over the different reservoirs that this state is in
contact with. The ability to grow the size of the N\'eel phase in this simple setup relies on the observation that the chemical potential of the N\'eel phase has a
dependence on the lattice depth which is well approximated by a power law.  Since this is likely to be the case for other 
phases of cold atoms in optical lattices we expect that the proposed setup will confer similar advantages to other attempts at realizing and probing such phases. One of the main challenges is
realizing a setup where the system is able to shed its entropy, even as a gapped phase is forming and inhibiting transport. Typically, present experiments rely on
precooling the atoms and then adiabatically loading them in to the lattice and forming the phase of interest.  Our proposed approach is to instead continue evaporative
cooling as the lattice is turned on, by maintaining the chemical potential at a level that allows the phase to stay in contact with a reservoir of mobile atoms that
is evaporatively cooled.






\acknowledgments
We would like to thank Pedro Duarte and Russell Hart for useful discussions.
C.J.M.M. acknowledges
support from the NSF through ITAMP at Harvard University and the Smithsonian
Astrophysical Observatory.
This work
was supported in part by ARO Award W911NF-07-1-0464 with
funds from the DARPA OLE Program. The work at Rice is also supported by the NSF, ONR, and the Welch Foundation (C-1133).

\bibliographystyle{apsrev4-1}

\end{document}